\documentclass[aps,twocolumn,superscriptaddress,preprintnumbers,showpacs,article,prl,longbibliography]{revtex4-2}

\usepackage[latin9]{inputenc}
\usepackage{color}
\usepackage{verbatim}
\usepackage{amsmath}
\usepackage{amssymb}
\usepackage{graphicx}
\usepackage{esint}
\usepackage{esvect}
\usepackage{braket}
\usepackage{amsmath,bm}
\usepackage{graphicx}
\usepackage{subfigure}
\usepackage{url}
\usepackage{lipsum}
\usepackage{lipsum}
\usepackage[dvipsnames]{xcolor}
\usepackage{hyperref}
\usepackage[resetlabels]{multibib}
\usepackage{ulem}

\makeatletter
\@ifundefined{textcolor}{}
{%
 \definecolor{BLACK}{gray}{0}
 \definecolor{WHITE}{gray}{1}
 \definecolor{RED}{rgb}{1,0,0}
 \definecolor{GREEN}{rgb}{0,1,0}
 \definecolor{BLUE}{rgb}{0,0,1}
 \definecolor{CYAN}{cmyk}{1,0,0,0}
 \definecolor{MAGENTA}{cmyk}{0,1,0,0}
 \definecolor{YELLOW}{cmyk}{0,0,1,0}
}

\usepackage{mathrsfs}

\draft
\makeatother
\begin{document}

\title{Fractional Chern insulator edges: crystalline effects and optical measurements}

\author{Yan-Qi Wang}
\affiliation{Joint Quantum Institute, University of Maryland, College Park, MD 20742, USA}

\author{Johannes Motruk}
\affiliation{Department of Quantum Matter Physics, University of Geneva, Quai Ernest-Ansermet 24, 1211 Geneva, Switzerland}

\author{Andrey Grankin}
\affiliation{Joint Quantum Institute, University of Maryland, College Park, MD 20742, USA}

\author{Mohammad Hafezi}
\affiliation{Joint Quantum Institute, University of Maryland, College Park, MD 20742, USA}

\begin{abstract}
Edge states of chiral topologically ordered phases are commonly described by chiral Luttinger liquids, effective theories that are exact only in the hydrodynamic limit. Motivated by recent bulk observations of fractional Chern insulators (FCIs) in two-dimensional materials and by synthetic realizations in ultracold atoms, we revisit this framework and quantify deviations from the hydrodynamic limit due to lattice effects. Using a combination of analytical arguments and numerical simulations, we disentangle universal from nonuniversal edge properties. We outline experimental probes in excitonic FCIs and in ultracold atom systems, and in particular propose time-resolved edge spectroscopy to directly access the predicted exponents and velocities.
\end{abstract}

\maketitle

%and can realize fractionalized states without magnetic fields when interactions exceed the bandwidth; they are being actively explored in van der Waals materials theoretically~\cite{abouelkomsan2020particle,wilhelm2021interplay,devakul2021magic,li2021spontaneous,crepel2023anomalous,wang2024fractional,reddy2023fractional,xu2024maximally,yu2023fractional,devakul2023magic,wang2023origin,ghorashi2023topological,Wang2023Breakdown} and experimentally~\cite{spanton2018observation,xie2021fractional,sharpe2019emergent,serlin2020intrinsic,zhang2019twisted,chen2020tunable,li2021quantum,foutty2024mapping,cai2023signatures,zeng2023integer,park2023observation,xu2023observation,lu2023fractional,makki2025probing,mostaan2025anyon}. 

{\it Introduction--} The fractional quantum Hall (FQH) effect arises when strong interactions in partially filled Landau levels produce incompressible liquids with fractional charge and anyonic statistics~\cite{Tsui_1982_PRB_TwoDimension}. Fractional Chern insulators (FCIs) are the lattice counterparts of FQH states. Originally, the bosonic FCI was proposed in Harper-Hofstadter settings at filling $\nu = 1/2$, defined by the ratio between particles and magnetic flux quanta~\cite{Kalmeyer1987Equivalence,HafeziPRA2007Fractional,Wang2022structure,Sorensen2005Fractional,Motruk2017Phase,Dong2018Charge}. The fermionic counterpart of such states has been extensively studied~\cite{regnault2011fractional,Neupert2011Fractional,Sun_2011_Fractional_NC}, and an anomalous version of these states, without external time-reversal breaking mechanism, has been recently observed in MoTe$_2$~\cite{cai2023signatures,zeng2023integer}. There have been significant efforts to implement bosonic FCIs in both ultracold atom systems~\cite{leonard2023realization} and superconducting qubit platforms~\cite{roushan2017chiral,rosen2024synthetic}. In both cases, however, the system size remains extremely small. Moreover, it has been predicted that excitonic versions of such states could emerge in semiconductor moir\'e heterostructures, where the topological nature of carriers can be inherited by the excitons~\cite{xie2024long}, and strong interactions between excitons can lead to the realization of FCIs~\cite{chen2025quantum}. Similar to cold atoms and superconducting qubits, excitons are charge neutral, and electronic transport measurements are therefore not relevant. Motivated by recent advances in probing strongly correlated systems via optical responses~\cite{Devereaux2023PRBAngleresolved,Gautum2023Correlation,Mcginley2022signatures,Fava2022divergent}, we ask whether optical probes can extract both universal and nonuniversal properties of FCIs in charge--neutral systems, such as excitons and cold atoms.

One central feature of topological phases is the bulk-edge correspondence~\cite{Alberto_2024_Quantum_PRL}: for example, a quantum Hall fluid hosts a chiral edge mode required by gauge invariance of the Chern--Simons effective theory on open manifolds. In the continuum limit, the edge is captured by a chiral Luttinger liquid ($\chi$LL), whose correlators exhibit power law decay and whose electron propagator has an anomalous exponent encoding the filling factor $\nu$. For one dimensional lattice models, the Luttinger description is asymptotic: band curvature generates controlled deviations from linear dispersion predictions~\cite{lukyanov2003long,pereira2008exact,PereiraPRB2009Spectral,Matveev2023Elementary}. The lattice thus introduces an additional energy scale in (fractional) Chern insulators, raising the question of how robust the $\chi$LL description is and which universal (e.g., exponents, topological data) and nonuniversal edge features can be reliably extracted from edge physics.

Here we show that, in the noninteracting limit, the lattice band curvature generically distorts equal--space edge correlators away from the hydrodynamic-$\chi$LL form. By contrast, interactions exponentially suppress these lattice corrections. We substantiate this claim with a combination of analytical (field-theory treatment and parton construction) and numerical (noninteracting lattice models and matrix product state simulations) approaches, and find remarkable qualitative agreement between them. While a substantial amount of work has shown commonalities between FQH states and FCIs in the bulk, our results highlight differences in the edge description. As an example, equal--space dynamical correlators show significant corrections to the power law scaling predictions~\cite{Wen1990PRBchiral}. Finally, we provide predictions for measurements of edge properties in terms of two point and density-density correlations, which allow extraction of parameters such as the filling and edge velocity in moir\'e materials, cold atom systems, and superconducting qubits.

% Moreover, we extract the anomalous boundary exponent, which tracks the bulk filling factor, from correlation functions, and independently determine the nonuniversal edge velocity and associated energy scales from short time dynamics. 

% \JM{forgot to say during the call, maybe flat band limit is not the best word? An extremely flat Chern band will still have the lattice corrections} 

% \JM{there are some redundancies now from copying in the abstract part, e.g. edge velocity mentioned twice}

\begin{figure}[!h]
\centering 
\includegraphics[width=1\columnwidth]{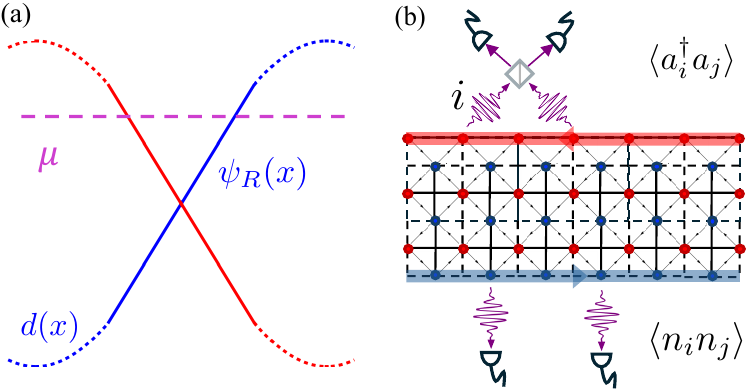}
\caption{\label{Figure_1}(a) Dispersion of the edge states on the strip. (b) Schematic illustration of the checkerboard lattice in a strip geometry and the measurement scheme. Black arrows and solid/dashed lines represent the nearest-neighbor (NN) and next-nearest-neighbor (NNN) hopping terms, respectively; the direction of each arrow indicates the sign of the complex phase in the NN hopping. Thick blue (red) arrows at the bottom (top) edge denote right- (left-) moving edge states. The measurement scheme, which allows extraction of the two point Green's function $\langle a_i^\dagger a_j\rangle$ and the density-density correlator $\langle n_i n_j\rangle$, is depicted schematically as two photodetectors. }
\end{figure}

{\it Field theory analysis--} According to the bulk-edge correspondence, the boundary of a Chern insulator supports gapless chiral edge states. If the edge dispersion is strictly linear, $\chi$LL theory predicts a pure power law decay of the correlation function in real space ($x$) and time ($t$), of the form $(x - v t)^{-m}$, where the exponent $m = 1/\nu$ reflects the filling factor $\nu$, and $v$ denotes the nonuniversal edge velocity~\cite{Wen1990PRBchiral}. However, a natural question arises: how is the correlation function modified when nonlinear effects such as band curvature are taken into account? For fermionic Chern insulators, we follow the analysis developed for conventional Luttinger liquids in strictly one dimensional systems~\cite{pereira2008exact,PereiraPRB2009Spectral} and calculate the equal--space correlation function for a $\chi$LL using bosonization. Without loss of generality, we focus on right-moving edge states, as illustrated in Fig.~\ref{Figure_1}(a). We begin by decomposing the edge fermion operator as $\psi(x) \sim e^{+i k_F x} \psi_R(x) + e^{+i k x} d(x)$, where $\psi_R(x)$ denotes the linearly dispersing component near $+k_F$, and $d(x)$ captures modes associated with curvature effects near the same momentum; see Fig.~\ref{Figure_1}(a) for a visual depiction. In the continuum limit, the noninteracting edge Hamiltonian density takes the form $H_{0} = -i v_F \psi_R^\dagger \partial_x \psi_R + d^\dagger (\epsilon_0 - i u_k \partial_x) d$, where $v_F$ is the bare linear velocity for $\psi_R$ at $k_F$, and $u_k$ is the bare velocity of the high energy particle. The bosonized expression for the right-moving fermion is given by $\psi_R(x) \sim e^{-i\sqrt{2\pi} \phi_R(x)}/\sqrt{2\pi \eta}$, where $\phi_R$ is the right moving part of the bosonic field and $\eta$ is a short distance cutoff (for example, the lattice constant). Then the bosonized noninteracting part of the edge Hamiltonian reads
\begin{equation}
    H_0 = \frac{v_F}{2} (\partial_x \phi_R)^2 + d^\dagger(\epsilon_0 - i u_k \partial_x)d.
\end{equation}
The density operator is given by
$n(x) = \psi^\dagger(x) \psi(x) \sim n_0 + \psi^\dagger_R \psi_R + d^\dagger d + e^{+i(k-k_F)x} \psi_R^\dagger d + e^{-i(k-k_F)x} d^\dagger \psi_R$.
We further assume that the interaction on the boundary is nearest neighbor, $H_{\rm int} \sim V \sum_i n_i n_{i+1}$.
One can then solve the full edge Hamiltonian $H_{\rm edge} = H_0 + H_{\rm int}$~\cite{Supp}, and define the single particle space time correlator
$G_h(x,t) = \langle \psi^\dagger(x,t) \psi(0,0)\rangle$.
Here and in what follows we denote the static correlator by
$G_h(x) = \langle \psi^\dagger(x,0) \psi(0,0)\rangle$
and the onsite time dependent Green's function by
$G_h(t) = \langle \psi^\dagger(0,t) \psi(0,0)\rangle$.
In the interaction picture, one can compute using field theory~\cite{pereira2008exact,PereiraPRB2009Spectral,Supp} that
\begin{equation}\label{eq:Correlation_Function}
    G_h(t) \sim \frac{A e^{-i{\mathcal E} t - t/\tau}}{t^\eta} + \frac{B}{t^\alpha}.
\end{equation}
The first term in Eq.~\eqref{eq:Correlation_Function} is the deviation from the standard $\chi$LL prediction and describes the transport of the nonlinear part of the edge states. Here, ${\mathcal E}$ is the energy difference between the chemical potential and the band bottom of the edge states, and this term decays with exponent $\eta = 0.5 + \delta \eta$, with $\delta \eta$ denoting the nonuniversal correction due to the interaction $V$. The decay time $\tau$ is finite and positive only for interacting systems ($V \neq 0$), so that in the long time limit one recovers the standard $\chi$LL behavior given by the second term; in the noninteracting limit, $\tau$ diverges~\cite{Supp}. The second term in Eq.~\eqref{eq:Correlation_Function} is the standard $\chi$LL prediction for a Chern insulator, with $\alpha = 1$~\cite{Wen1990PRBchiral}. Due to the superposition of two different scalings, the absolute value of $G_h(t)$ exhibits an oscillating behavior in time, with periodicity determined by ${\mathcal E}$.

For fractional Chern insulators (FCIs), it is difficult to construct an effective edge theory directly in terms of the original particles, as such descriptions fail to capture the anyonic excitations present at the boundary. To overcome this, we introduce a parton construction, where the annihilation operator for an elementary particle at site $i$ is expressed as $a_i = \epsilon_{\lambda_1 \cdots \lambda_m} f_{i\lambda_1} \cdots f_{i\lambda_m}$, with $f_{i\lambda}$ denoting the annihilation operator for a fermionic parton with $m$ colors (labeled by $\lambda$). For the bosonic case at $\nu = 1/2$, we have $a_i = f_{i1} f_{i2}$~\cite{PichlerPRB2025Single,LuPRB2012Symmetry,McGreevy2012Wave}. At the mean field level, the Green's function in real space for the elementary particle along the edge is the product of two parton Green's functions:
\begin{equation}\label{eq:Parton_Correlation_Function}
    \Gamma_h(t) = \langle a_0^\dagger(t) a_0(0) \rangle = \prod_{\lambda=1}^2 G_h^\lambda(t),
\end{equation}
where each $G_h^\lambda(t)$ is given by Eq.~\eqref{eq:Correlation_Function} with nonzero $\tau$. In the long time limit, ignoring the oscillating terms, the factor $e^{-t/\tau}/t$ is exponentially suppressed (but survives in the noninteracting system and dominates at long times), and the product of the remaining $1/t$ contributions yields $\Gamma_h(t) \sim 1/t^2$, in accordance with hydrodynamic theory~\cite{Wen1990PRBchiral}, which predicts $\langle a_x^\dagger(t) a_0(0) \rangle \sim 1/(x - v t)^{1/\nu}$. The parton construction is valid when the edge is fully chiral: all parton modes propagate in the same direction, so the gauge constraint is approximately satisfied by the locking of the currents, which gaps out neutral channels. What remains is a single charged chiral mode with universal Luttinger scaling, making the parton construction a faithful and natural description of FCI edges, and often more controlled than in the bulk.

\begin{figure}[!h]
\centering 
\includegraphics[width=1\columnwidth]{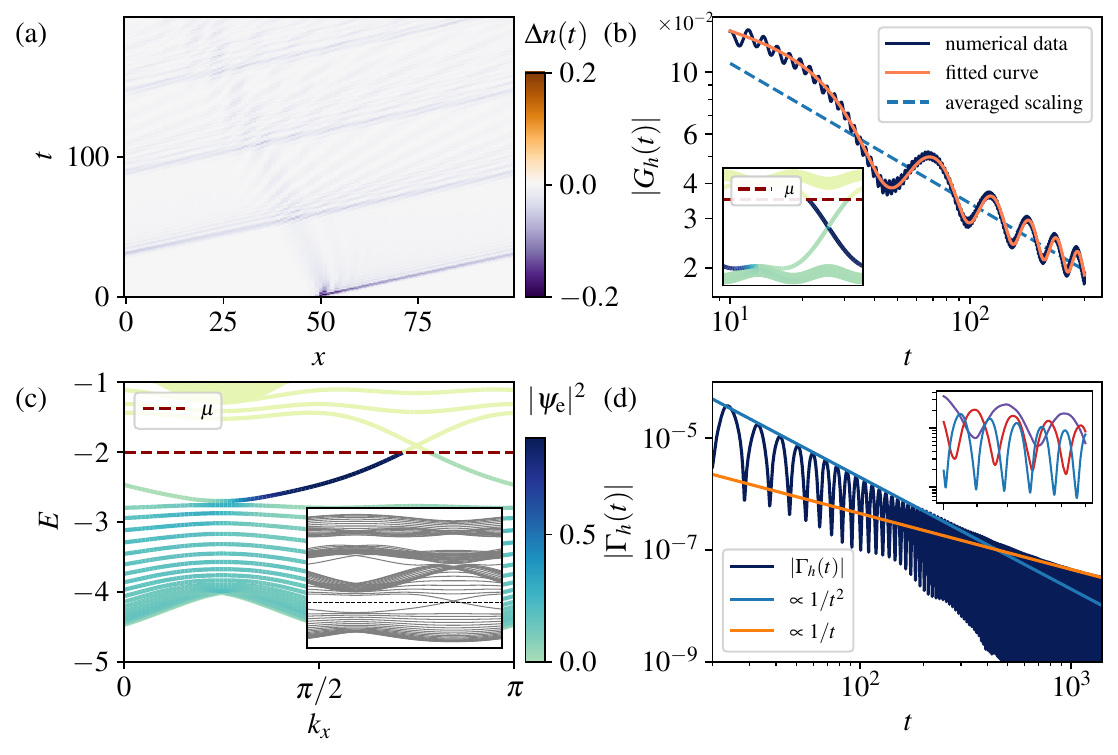}
\caption{\label{Figure_Free_Dynamic}
Dynamics of Chern insulators and parton FCIs.
(a) Boundary charge density difference $\Delta n(t)$ as a function of time for a Chern insulator with
$t_1 = 1$, $t_{2a} = -t_{2b} = 1/\sqrt{2}$, $\phi = -\pi/4$, and $\mu = 1.4 t_1$ on a strip of size
$L_x = 100$, $L_y = 50$.
(b) Dynamical correlator $G_h(t)$ for the same parameters on a larger strip, $L_x = 600$, $L_y = 120$.
A fit to Eq.~\eqref{eq:Correlation_Function} yields $\eta = 0.50$, $\alpha = 0.96$, and
${\mathcal E} = 0.12$.
Inset: single particle band structure with color indicating boundary weight (darker means larger edge amplitude).
(c) Lower part of the parton band structure for the Hamiltonian in Eq.~(4) on a strip with
$L_x = 600$, $L_y = 100$, $\tilde t_1 = t_1$, $\tilde t_{2a} = \sqrt{t_{2a}}$, $\tilde t_{2b} = i \tilde t_{2a}$,
and $\mu = -2.0 \tilde t_1$.
The color scale indicates the parton weight on the boundary; the inset shows the full band structure.
(d) Onsite dynamical parton correlator $\Gamma_h(t)$ from Eq.~\eqref{eq:Parton_Correlation_Function}.
In the absence of interactions, the correlator initially follows a $t^{-2}$ decay set by the linear edge dispersion (blue), and crosses over to $t^{-1}$ at longer times as band curvature becomes relevant (orange).
Inset: $\Gamma_h(t)$ for the same system size at $\mu = -2.4$ (purple), $-2.2$ (red), and $-2.0$ (blue).
Note that the oscillation frequency increases with increasing energy difference between $\mu$ and the band bottom of the edge dispersion.
}
\end{figure}

{\it Lattice Model--} We consider a two-band model defined on the checkerboard lattice~\cite{Sun_2011_Nearly_PRL,Sun_2011_Fractional_NC,Neupert2011Fractional} with nearest-neighbor $\langle i,j\rangle$ hopping $t_1$ and next-nearest-neighbor $\langle\langle i,j\rangle \rangle$ hoppings $t_{2a}$ and $t_{2b}$. Here, we allow the nearest-neighbor hopping to carry a nonzero complex phase $(\pm \phi)$, whose signs are shown by the arrows; see Fig.~\ref{Figure_1}(b) for further details. These complex hoppings break time reversal symmetry at $\phi \neq n \pi$ ($n \in {\mathbb Z}$) and lead to two bands with Chern numbers $C = \pm 1$. The Hamiltonian of this model reads
\begin{equation}\label{eq:Lattice_Hamiltonian}
    \begin{aligned}
        H =& - t_1 \sum_{\langle i,j \rangle} e^{i\phi_{ij}} a_i^\dagger a_j
             - \sum_{\langle \langle i,j \rangle \rangle} t_{2}^{ij} a_i^\dagger a_j + {\rm h.c.}.
    \end{aligned}
\end{equation}
The operators $a_i$ and $a_i^\dagger$ commute (anticommute) if they represent hard-core bosons (fermions). The hopping strength between NNN sites takes the value $t_{2a}$ ($t_{2b}$) if the two sites are connected by a solid (dashed) line.

{\it Time evolution--} In this section, we demonstrate how to numerically extract both universal (e.g., filling factor) and nonuniversal (e.g., edge velocity) properties from the lattice model in Eq.~\eqref{eq:Lattice_Hamiltonian}. We implement the model Hamiltonian on a strip geometry with periodic boundary conditions along the $x$-direction and open boundary conditions along the $y$-direction. To probe the edge dynamics, we initialize the system by annihilating a particle at the center of the strip and compute the difference in charge density $\Delta n(t) = \langle n(t) \rangle - n_0$ along the boundary (see Fig.~\ref{Figure_Free_Dynamic}(a)). The resulting edge excitation propagates ballistically along the boundary in the negative $x$-direction, with the slope of the trajectory indicating an edge velocity $v_F \approx 5/3~(t_1 a/\hbar)$, where $a$ is the lattice constant. This agrees with the slope of the inset in Fig.~\ref{Figure_Free_Dynamic}(b), which yields the edge velocity from the single particle energy spectrum. We numerically compute the onsite dynamical single particle correlation function $\langle a_i^\dagger(t)\, a_i(0) \rangle$, shown as the blue line in Fig.~\ref{Figure_Free_Dynamic}(b). The scaling behavior matches the prediction from Eq.~\eqref{eq:Correlation_Function}.

When $a_i^\dagger$ denotes the creation operator for a hard-core boson at site $i$, we numerically confirm that the system realizes a bosonic fractional Chern insulator (FCI) at filling $\nu = 1/2$~\cite{Supp}. To study the time evolution, we begin by analyzing the dynamics of the parton model. The parton construction for the model Hamiltonian was introduced in Refs.~\cite{McGreevy2012Wave,LuPRB2012Symmetry}, and we briefly review it below. In this approach, each sublattice of the checkerboard lattice is divided into two parton flavors, with each parton experiencing the square root of the original hopping amplitude. When combined, the partons reproduce the full flux and hopping structure of the original bosonic operators. Figure~\ref{Figure_Free_Dynamic}(c) shows the band structure of the partonized version of Eq.~\eqref{eq:Lattice_Hamiltonian}. We numerically compute the onsite dynamical correlation functions $G_h^\lambda(t)$ for each parton flavor. Since the parton model in principle includes an interaction term, the first (exponentially decaying) term in Eq.~\eqref{eq:Correlation_Function} should be suppressed in the long time limit. In our mean-field treatment, we take the product of the single-parton correlation functions of each flavor, which yields the correlation function shown in Fig.~\ref{Figure_Free_Dynamic}(d). Here, the correct long time behavior $\propto 1/t^2$ is reproduced; however, it reverts back to $1/t$ at very long times due to the missing exponential suppression  of the corrections from curvature, as in the integer Chern insulator.

\begin{figure}[!h]
\centering 
\includegraphics[width=1\columnwidth]{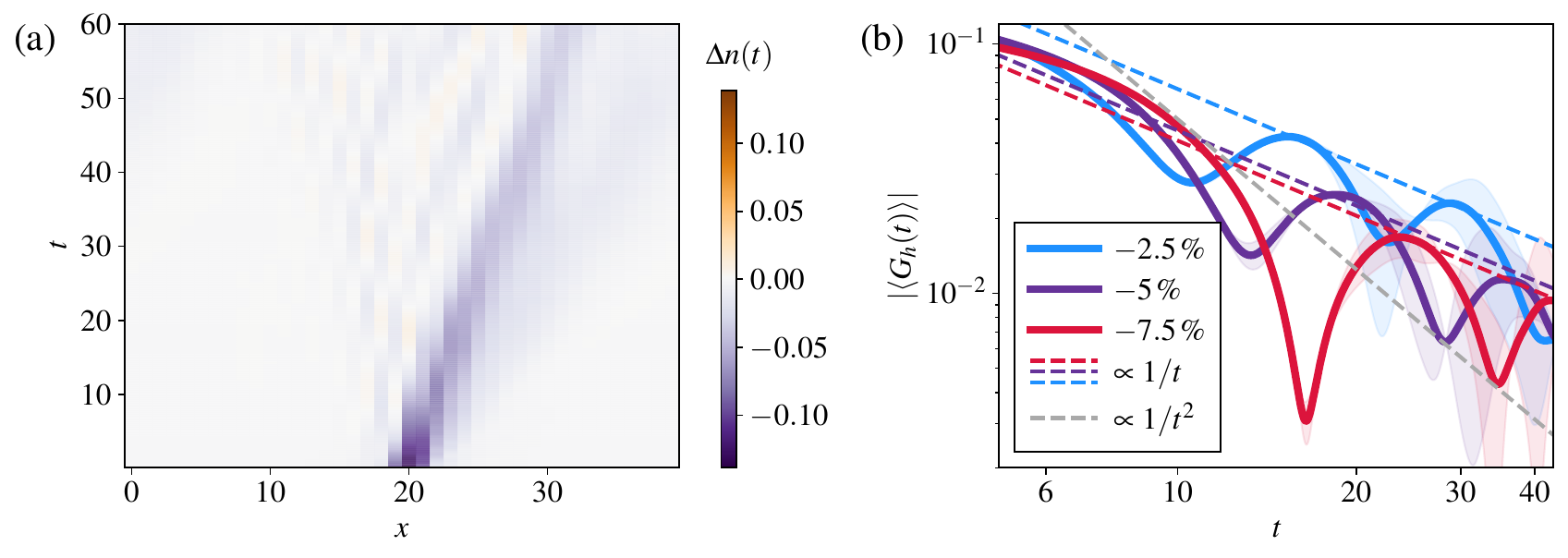}
\caption{\label{Figure_tDMRG}
Dynamics of the bosonic FCI. The particle filling factor is defined by
$\nu_b = N_b/(2 L_x L_y)$, corresponding to a lowest band filling of
$\nu = 1/2$ with periodic boundary conditions. Time evolution is computed
on a strip of $L_x = 40$, $L_y = 4$ with periodic structure along the
$x$-direction. In a strip with finite $L_y$, a lower density $\nu_b$ is
used to account for the reduced particle density at the edges formed by
the hard confinement~\cite{Supp}. (a) Boundary charge density as a
function of time for a $-7.5\%$ deviation of the particle number from
$\nu_b = 1/4$. (b) Dynamical onsite single particle correlator for
several fillings (deviations from $\nu_b = 1/4$) for bond dimension
$\chi = 400$; the bulk particle density remains unchanged for these
cases~\cite{Supp}. The shaded regions indicate error margins determined
by the difference between the $\chi = 400$ and $\chi = 200$ results,
which provides a conservative error estimate.
}
\end{figure}

To study the time evolution for the bosonic FCI beyond the parton construction, we consider the full many-body model on an infinite strip. We compute the ground state $\ket{\Psi_G}$ with DMRG and then simulate the time evolution of $\ket{\Psi_M} = a_0 \ket{\Psi_G}$ under the Hamiltonian $H$ using a matrix product operator based time evolution method~\cite{Zaletel2015}. Here, $\ket{\Psi_M(t)}$ is the wave function at time $t$, and $a_0$ annihilates a boson at the bottom edge of the strip. In particular, we use an infinite MPS (iMPS) unit cell of length $L_x=40$ and width $L_y=4$ unit cells (8 sites). We define the difference in charge density at time $t$ as $\Delta n(t) = \bra{\Psi_M(t)} a^\dagger_{i} a_{i} \ket{\Psi_M(t)} - \bra{\Psi_G} a^\dagger_{i} a_{i} \ket{\Psi_G}$,
and plot it at the bottom edge as a function of time. The wave front of the edge charge density spreads ballistically, and its slope denotes the edge velocity; see Fig.~\ref{Figure_tDMRG}(a). We then compute the onsite dynamical correlation function $G_h(t) = \bra{\Psi_G} a_{0}^\dagger(t) a_{0}(0) \ket{\Psi_G}$. As shown in Fig.~\ref{Figure_tDMRG}(b), $G_h(t)$ clearly deviates from the asymptotic $1/t^2$ decay (dashed line) predicted by Wen~\cite{Wen1990PRBchiral} in the short time window that we consider. Instead, we observe a scaling dominated by $1/t$ (dashed lines). This arises from the modification by the crystalline structure, in accordance with the discussion in the analytic sections. The oscillating behavior of the absolute value of $G_h(t)$ reflects the superposition of two scalings. The oscillation frequency changes with particle number (filling), resulting in slower oscillations for lower filling, in qualitative agreement with the parton mean-field theory. If the detection time of $G_h(t)$ in realistic platforms is limited to short times due to decoherence, this $1/t$-dominated regime will be the experimental observation.

\begin{figure}[!h]
\centering 
\includegraphics[width=1\columnwidth]{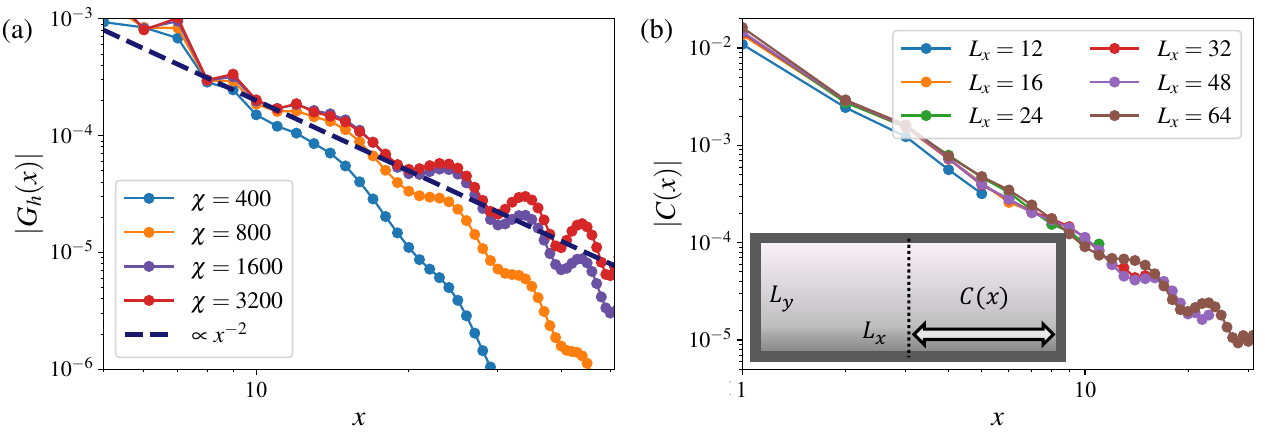}
\caption{\label{Figure_Static_Correlation}
Static correlators on a strip of width $L_y = 4$ with a filling deviation of $-5\%$.
(a) Infinite strip along $x$ (iDMRG) with bond dimensions
$\chi = 400, 800, 1600, 3200$ for the single particle correlator
$|G_h(x)|$.
(b) Finite  strips of lengths $L_x = 12, 16, 24, 32, 48, 64$ at fixed
$\chi = 1600$ for the connected density--density correlator $|C(x)|$.
In both panels, $x$ measures the distance along the lower edge; in (b),
$C(x)$ is measured from the center of the lower edge, as indicated in the inset.}
\end{figure}

{\it Static correlators--} Complementing the time-domain analysis on the infinite strip, we now turn to static equal-time correlators, which provide an independent benchmark of the ground-state edge theory. In Fig.~\ref{Figure_Static_Correlation}(a), we display $G_h(x,0)$, which shows the characteristic decay of $\sim 1/x^{1/\nu}$. We also compute the connected density--density correlation function
$C(x)\equiv \langle n_x n_0\rangle - \langle n_x\rangle\langle n_0\rangle$, shown for finite size strips in Fig.~\ref{Figure_Static_Correlation}(b), which exhibits a clear power law decay consistent with the bosonization prediction, being proportional to $1/\!\bigl[(4\pi^2)(x+ i a)^2\bigr]$.
%, as shown in Fig.~\ref{Figure_Static_Correlation}(a). The mild deviations at large distances arise from the finite bond dimension and the finite transverse width of the strip. The same power law behavior is also observed in finite systems, as displayed in Fig.~\ref{Figure_Static_Correlation}(b).

Our results are relevant to experimental efforts to observe bosonic FCIs in moir\'e materials, ultracold atoms in optical lattices, and superconducting qubits. Below, we discuss measurement protocols for the first two platforms.

\textit{Moir\'e systems--} Moir\'e superlattices in two-dimensional TMD semiconductors provide a highly controllable platform where strong electronic correlations and nontrivial topology intersect. Specifically, the moir\'e pattern strongly localizes excitons, leading to the observation of a correlated incompressible excitonic state -- a bosonic Mott insulator -- described by a Bose-Hubbard-type model on a triangular lattice 
\cite{xiong2023correlated,park2023dipole,gao2024excitonic,lian2024valley,miao2021strong,xiong2024tunable}. These findings raise the exciting possibility of engineering topological moir\'e bands for excitons, which could host bosonic correlated topological phases such as bosonic fractional Chern insulators~\cite{perea2025exciton,xie2024long,froese2025topological}. Based on this, we now describe an optical measurement scheme that enables the detection of correlation functions at the edge of a bosonic FCI in an excitonic system. The setup we have in mind is shown in Fig.~\ref{Figure_1}(b). Specifically, we consider an optically pumped excitonic state on a lattice, interacting via contact two-body interactions described by Eq.~\eqref{eq:Lattice_Hamiltonian}~\cite{xie2024long,chen2025quantum}. We further assume that the excitons decay slowly by emitting photons, with the corresponding dissipation rate being smaller than the characteristic energy scales of the system governed by Eq.~\eqref{eq:Lattice_Hamiltonian}. This dissipation rate thus sets an upper bound on the longest time that can be measured, which is directly relevant for the time window shown in Fig.~\ref{Figure_tDMRG}(b). The emission from different lattice sites can be collected and processed using standard tools of linear optics, enabling both momentum and frequency resolution of the correlation functions. Assuming the bulk is gapped, the emission from the system's edge can be filtered. The relevant correlation function describing the emission properties in a frame rotating at the excitonic resonance frequency is given by the two-point function~\cite{walls_milburn_quantum_optics}, as shown in Fig.~\ref{Figure_1}(b),
\begin{equation}
G_{i,j}^{(1)}(t)=\langle a_{j}^{\dagger}(t)a_{i}(0)\rangle, \label{eq:G1}
\end{equation}
where $a_x^\dagger(t)\equiv e^{iHt}a_x^\dagger e^{-iHt}$. We note that far-field radiation does not allow one to resolve sites separated by distances smaller than the wavelength of the emitted light. This limitation can be overcome by employing near-field techniques, such as scattering-type scanning near-field optical microscopy (s-SNOM)~\cite{hillenbrand2025visible}, which provides spatial resolution on the order of 10~nm. Such resolution is comparable to the lattice spacing in moir\'e TMD systems, enabling measurements of $G^{(1)}_{i,j}(t)$. For $\nu=1/2$ bosonic FCIs, the universal data, such as the filling factor, can be extracted either from the exponent of the static spatial scaling of $G_{i,j}^{(1)}(0)$ or from the onsite temporal scaling of $G_{i,i}^{(1)}(t)$. For example, the static correlator $G_{0,x}^{(1)}(0)$ exhibits a $1/x^2$ scaling, as shown in Fig.~\ref{Figure_Static_Correlation}(a). Furthermore, the same setup allows for the measurement of the four-point (density-density) correlation function~\cite{walls_milburn_quantum_optics}. The non-universal data, such as the edge velocity, can be obtained by measuring $\langle a_{i}^{\dagger}(t)a_{i}(t)\rangle$ at the same site $i$, which yields a time-resolved edge-density profile, as shown in Fig.~\ref{Figure_Free_Dynamic}(a) or Fig.~\ref{Figure_tDMRG}(a).

% Two-point onsite correlators follow our prediction presented in Fig XX and XXX. Two-point equal-time correlators are presented in the Fig XXX.

%d

%\begin{figure}[!h]
%\centering 
%\includegraphics[width=1\columnwidth]{Figure_main_5_measurement.pdf}
%\caption{\label{Figure_Measurement} \JM{change font of (a) and (b) to Times} Schematic illustration of the measurement scheme. {(a) Solid-state moir\'e TMD measurement scheme.  Light emitted due to the exciton recombination from the edge sites $i$ and $j$ is collected and processed using linear optical elements, represented here as a beam splitter (shown as a gray square).  $G_{i,j}^{(1)}(t=0)$ (Eq.~\eqref{eq:G1}) can be measured by subtracting the photon fluxes in the two photodetectors. The equal-time four-point correlation function can be extracted by measuring the coincidences in photon arrival at the detectors\cite{walls_milburn_quantum_optics}. (b) Cold-atom setup: the light is scattered on the atoms in an optical lattice and then detected using a quantum gas microscope (shown schematically as two detectors). This measurement provides the information on the atomic positions thereby allowing for the extraction of the four-point equal-time density-density correlation function $\langle n_i n_j\rangle$. } \JM{where is the tip in part (b)?}
%}\ag{yes after discussing now keep it at conceptual level  {\YQW{also, would it be better to unify the plots style for (a),(b)?}} }
%\end{figure}

{\it Ultracold atoms in optical lattices--} Early proposals to detect bosonic FCIs focused on collective properties, such as incompressibility and the magneto-roton minimum~\cite{HafeziPRA2007Fractional}, and more recently on bulk properties such as Hall drift measurements~\cite{Motruk2020,repellin2020fractional}. The advent of the quantum gas microscope provides the ability to measure various types of site-resolved density-density correlations~\cite{bakr2009quantum}. In contrast to the exciton case discussed above, measuring coherent correlations between two remote sites remains challenging in cold atom systems~\cite{impertro2024local}. Therefore, the static density-density correlators shown in Fig.~\ref{Figure_Static_Correlation}(b) are more relevant. Based on Fig.~\ref{Figure_Static_Correlation}(b), the edge correlators remain meaningful even in small lattices (see, e.g., Ref.~\cite{leonard2023realization}), making such measurements a promising scheme to demonstrate the chiral and 1D nature of topological edge states.

%In summary, we analyzed the deviation from standard power-law scaling at the boundary of a non-interacting Chern insulator due to band curvature, using both field-theoretical methods and numerical simulations on a lattice model. We showed that this deviation is exponentially suppressed in interacting systems, such as those in the fractional Chern insulator (FCI) phase. We further computed the spatial and temporal edge correlation functions for a bosonic FCI phase using MPS methods and parton construction, respectively. Finally, we proposed methods to measure these correlation functions in bosonic systems through their optical response, enabling the extraction of both universal data (e.g., filling fraction) and non-universal parameters (e.g., edge velocity). 
{\it Outlook--} As future directions, it would be valuable to study the real time quantum dynamics using newly developed methods such as isometric tensor networks~\cite{Dai2025Fermionic}, and to investigate optical probes for other topological orders, such as chiral superconductors or the $\nu = 2/3$ FQHE state~\cite{wang2024chiral,tam2025quantized,Devereaux2023PRBAngleresolved}.

{\it Acknowledgements--} We acknowledge insightful discussions with Joel E. Moore, Jay D. Sau, Biao Lian, Pok Man Tam, Thierry Giamarchi, Michael Knap, Monika Aidelsburger, Louk Rademaker, and Zhi-Qiang Gao. Y.-Q. W. is supported by the JQI postdoctoral fellowship at the University of Maryland. J.M. was supported by the SNSF Swiss Postdoctoral Fellowship grant No.~210478. MPS simulations have been performed using the \texttt{TeNPy} library~\cite{tenpy} and one of its previous versions~\cite{Kjall2013} on the Baobab and Mafalda HPC clusters at the University of Geneva.

{\it Note added.} While preparing this draft, we became aware of a recent study that reported the long time and long distance edge response using a different MPS method~\cite{yang2025}.

% While preparing this draft, we became aware of a recent numerical study~\cite{yang2025} which shows, complementary to our short-time analysis, that the long-time/long-distance edge response reverts to the universal prediction. 

% While preparing this draft, we became aware of a recent study which shows, using a different MPS method, that the long-time/long-distance edge response has been reported~\cite{yang2025}.

% reverts to the universal prediction.

\bibliographystyle{apsrev4-2}
\bibliography{apssamp}

\noindent

\onecolumngrid

\renewcommand\theequation{S\arabic{equation}}
\renewcommand\thefigure{S\arabic{figure}}
\renewcommand\bibnumfmt[1]{[S#1]}
\setcounter{equation}{0}
\setcounter{figure}{0}
\setcounter{secnumdepth}{3}

\clearpage{}
\section*{Supplementary Material}

\section{Bosonization theory for crystalline modification to chiral edges}\label{Supp_Sec:chiLL}

In this section, we calculate how the crystalline structure (i.e., band curvature) modifies the chiral Luttinger liquid as the boundary theory of an interacting Chern insulator, following Refs.~\cite{pereira2008exact,PereiraPRB2009Spectral}. We begin by decomposing the right-moving edge fermion operator as 
\begin{equation}\label{Appendix:Eq:Expansion}
    \psi(x) \sim e^{ik_F x} \psi_R(x) + e^{ik x} d(x),
\end{equation}
where $\psi_R(x)$ denotes the linearly dispersing component near the Fermi momentum $+k_F$, and $d(x)$ captures the modes associated with curvature effects in the nearby momentum range $k_F < k < k_M$, as illustrated in Fig.~\ref{Appendix_Figure_1}. In the continuum limit, the non-interacting edge Hamiltonian density takes the form
\begin{equation}\label{Appendix:Eq:Hamiltonian}
    H_0 = -i v_F \psi_R^\dagger \partial_x \psi_R + d^\dagger(\epsilon_0 - i u_k \partial_x) d,
\end{equation}
where $v_F$ is the bare linear velocity for $\psi_R$ at $k_F$, $\epsilon_0$ is the linearized free-fermion dispersion $\epsilon^{(0)}(k)$, and $u_k$ is the momentum-dependent bare velocity of the high-energy particle. One example is a trigonometric modification to the linear dispersion, such as $\epsilon^{(0)}(k) = -2\cos k - \mu$ for $0 < k_F < k < \pi = k_M$. The ground state is constructed by filling single-particle states up to the Fermi level, $\epsilon^{(0)}(k_F) = 0$. The bosonized expression for the right-moving fermion is given by
\begin{equation}\label{Appendix:Eq:Bosonized_Field}
    \psi_R(x) \sim \frac{1}{\sqrt{2 \pi \eta}} e^{-i\sqrt{2\pi} \phi_R(x)},
\end{equation}
where $\phi_R$ is the right-moving part of a bosonic field and $\eta$ is a short-distance cutoff (for instance, the lattice constant). With this, Eq.~\ref{Appendix:Eq:Hamiltonian} can be bosonized as
\begin{equation}\label{Appendix:Eq:Bosonized_Free_Hamiltonian}
    H_0 = \frac{v_F}{2} (\partial_x \phi_R)^2 + d^\dagger(\epsilon_0 - i u_k \partial_x) d.
\end{equation}

The interaction part, together with the chemical potential term, at the lattice Hamiltonian level after projecting onto the edges, reads
\begin{equation}
    H_{\rm int} = \sum_{j=1}^L \bigg[ V \bigg( n_j - \frac{1}{2} \bigg) \bigg( n_{j+1} - \frac{1}{2} \bigg) - \mu n_j \bigg].
\end{equation}
Here, $L$ is the total number of sites along the $x$-direction of one edge, $V > 0$ is the projected strength of the repulsive interaction, $n_j = \psi_j^\dagger \psi_j$ is the number operator at site $j$, and $\mu$ is the chemical potential. Assuming translational invariance, we define $n \equiv \langle n_j \rangle$, and, at half-filling, $n = 1/2$. Expressing the density operator $n_x$ with the help of Eq.~\ref{Appendix:Eq:Expansion}, we obtain its low-energy form,
\begin{equation}\label{Appendix:Eq:n(x)}
    n_x \sim n + \psi_R^\dagger \psi_R + d^\dagger d + e^{+i(k-k_F)x} \psi_R^\dagger d + e^{-i(k-k_F)x} d^\dagger \psi_R.
\end{equation}

Substituting Eq.~\ref{Appendix:Eq:n(x)} into $H_{\rm int}$ and using the bosonization relation Eq.~\ref{Appendix:Eq:Bosonized_Field}, we find the Hamiltonian density for the interaction term as
\begin{equation}\label{Appendix:Eq:Int_Hamiltonian}
    \begin{aligned}
        H_{\rm int} \sim \frac{V \sin^2 k_F}{\pi} (\partial_x \phi_R)^2 
        - \frac{2V}{\pi} \sin k_F \big( \cos k\, d^\dagger d + i \sin k\, d^\dagger \partial_x d \big)
        - \frac{4V \sin^2[(k-k_F)/2]}{\sqrt{2\pi}} (\partial_x \phi_R) d^\dagger d,
    \end{aligned}
\end{equation}
where we have set $\eta = 1$ and omitted irrelevant interaction terms and the renormalization of the chemical potential. 

By combining the bosonized free Hamiltonian Eq.~\ref{Appendix:Eq:Bosonized_Free_Hamiltonian} and the bosonized interacting Hamiltonian Eq.~\ref{Appendix:Eq:Int_Hamiltonian}, we obtain
\begin{equation}
    \begin{aligned}
        H &= H_0 + H_{\rm int} 
        = \frac{v}{2} (\partial_x \phi_R)^2 + d^\dagger(\epsilon - i u \partial_x) d 
        - \frac{1}{\sqrt{2\pi K}} \kappa_R (\partial_x \phi_R) d^\dagger d,
    \end{aligned}
\end{equation}
where we define the following renormalized parameters:
\begin{equation}
    \frac{v}{2} = \frac{v_F}{2} + \frac{V \sin^2 k_F}{\pi}, \quad
    \epsilon = \epsilon_0 - \frac{2V}{\pi} \sin k_F \cos k, \quad
    u = u_0 + \frac{2V \sin k_F}{\pi} \sin k, \quad
    \frac{\kappa_R}{\sqrt{2\pi K}} = \frac{4V \sin^2[(k-k_F)/2]}{\sqrt{2\pi}},
\end{equation}
with $K = 1$ for a non-interacting Chern insulator. 

We can decouple the $d$ particle from the bosonic fields by performing a unitary transformation
\begin{equation}
    U = \exp \bigg[ - \frac{i}{\sqrt{2\pi K}} \int_{-\infty}^{+\infty} dx\, (\gamma_R \phi_R) d^\dagger d \bigg],
\end{equation}
such that $\bar \phi_R = U \phi_R U^\dagger$, with
\begin{equation}
    \gamma_R(k) = \frac{\kappa_R(k)}{v - u(k)}.
\end{equation}
In terms of the transformed fields $\bar \phi_R = U \phi_R U^\dagger$ and $\bar d = U d U^\dagger$, we have
\begin{equation}
    \partial_x \phi_R = \partial_x \bar \phi_R + \frac{\gamma_R}{\sqrt{2\pi K}} \bar d^\dagger \bar d, 
    \quad
    d = \bar d \exp \bigg[ -\frac{i}{\sqrt{2\pi K}} (\gamma_R \bar \phi_R) \bigg].
\end{equation}
The Hamiltonian density then becomes non-interacting up to irrelevant operators:
\begin{equation}
    \mathcal{H} = \frac{v}{2} (\partial_x \bar \phi_R)^2 + \bar d^\dagger(\epsilon - i u \partial_x) \bar d.
\end{equation}

\subsection{Particle Green's function}
We are now ready to compute the time-dependent particle Green's function,
\begin{equation}
    G_p(x,t) = \langle \psi(x,t)\psi^\dagger(0,0) \rangle = G_p^R (x,t) + G_p^d(x,t),
\end{equation}
where $G_p^R(x,t) = \langle \psi_R(x,t) \psi^\dagger_R(0,0) \rangle$ and $G_p^d(x,t) = \langle d(x,t) d^\dagger(0,0) \rangle$. In the long-time limit $t \gg \epsilon^{-1}$, we can employ the mode expansion and write
\begin{equation}
    G_p^d(k,t) \sim \int dx\, \langle d(x,t) d^\dagger(0,0) \rangle.
\end{equation}
Expressed in terms of the decoupled fields, we have
\begin{equation}
    \begin{aligned}
        G_p^d(k,t) \sim \int dx\, \langle \bar d(x,t) \bar d^\dagger(0,0) \rangle \,
        \langle e^{-i\sqrt{2\pi \nu_R}\bar \varphi_R(x,t)} e^{+i \sqrt{2\pi \nu_R} \bar \varphi_R(0,0)} \rangle,
    \end{aligned}
\end{equation}
where 
\begin{equation}
    \nu_R(k) = \frac{1}{K} \bigg( \frac{\gamma_{R}(k)}{2\pi} \bigg)^2
\end{equation}
is the anomalous exponent. The free propagator for the $d$ particle takes the form
\begin{equation}
    G_p^{d,(0)}(x,t) = \langle \bar d(x,t) \bar d^\dagger(0,0) \rangle = e^{-i\epsilon t}\delta(x - u t).
\end{equation}
The correlation function of the free bosonic fields can be evaluated by standard methods,
\begin{equation}
    \langle e^{-i\sqrt{2\pi \nu_R}\bar \varphi_R(x,t)} e^{+i \sqrt{2\pi \nu_R} \bar \varphi_R(0,0)} \rangle \sim \frac{1}{t^{\nu_R}},
\end{equation}
yielding the asymptotic long-time behavior of the correlation function,
\begin{equation}
    G_p^d(k,t) \sim \frac{e^{-i{\mathcal E} t}}{t^{\eta_R}},
\end{equation}
with $\eta_R = \nu_R(k)$.

We now analyze the long-time behavior of the fermionic Green's function in real space, which is dominated by the saddle-point contribution. For the particle Green's function, the relevant saddle corresponds to a particle at the bottom of the band, $k = 0$; for the hole contribution, it corresponds to a particle at the top of the band, $k = \pi$. The high-energy excitations near these points have a parabolic dispersion, $\epsilon(k) \approx \epsilon + k^2/2m$, where $m < 0$ is the renormalized mass. Consequently, the propagator for the decoupled $\bar d$ particle reads (see Sec.~\ref{Supp_Sec:Mode_Expansion} for further details)
\begin{equation}\label{eq:supp:particle_bare_scaling}
    G_p^{d,(0)}(x,t) \sim \sqrt{\frac{m}{2\pi i t}} e^{-i\epsilon_\pi t - i m x^2 /2t}.  
\end{equation}
Therefore, in the non-interacting case, this contribution to the particle Green's function oscillates at a high frequency $\epsilon_\pi \sim O(1)$ for $t \gg |m|x^2$ and decays as $1/\sqrt{t}$. In the interacting case, the oscillation frequency becomes the renormalized energy of the single particle with momentum $k = \pi$. In addition, the power-law exponent is modified by coupling to the low-energy modes. For the particle Green's function below half-filling, there is also an exponential decay associated with the decay rate $1/\tau_\pi$ of the particle at $k = \pi$.

We consider a quasiparticle with momentum $k$ and energy $\omega$ above the Fermi sea $\ket{\rm GS}$, with the initial state $\ket{i} = \psi_k^\dagger\ket{\rm GS}$. To order $V^2$, the only inelastic decay channel is the creation of a particle-hole pair in the sea, with the final state $\ket{f} = \psi^\dagger_{k-q}\psi^\dagger_{p+q} \psi_p \ket{\rm GS}$, subject to the constraints $n_p = 1$, $n_{p+q} = 0$, and $n_{k-q} = 0$ ($\theta(-\epsilon_p^{(0)}) \theta(\epsilon_{p+q}^{(0)}) \theta(\epsilon_{k-q}^{(0)})$). Energy conservation for the transition gives
\begin{equation}
    E_f - E_i = \epsilon^{(0)}_{k-q} + \epsilon^{(0)}_{p+q} - \epsilon^{(0)}_p - \omega.
\end{equation}
Details are illustrated in Fig.~\ref{Appendix_Figure_1}(a). The matrix element $M_{fi} = \langle f | H_{\rm int} | i \rangle = \frac{1}{2L}[V(q) - V(k - p - q)]$ has two contributions (direct and exchange). Fermi's golden rule gives the retarded self-energy
\begin{equation}
        {\rm Im} \Sigma^{(2)}(k,\omega) = -\frac{\pi}{2} [V(q) - V(k - p - q)]^2 \theta(\epsilon_{k-q}^{(0)})\theta(-\epsilon^{(0)}_p) \theta(\epsilon^{(0)}_{p+q}) \delta(\omega - \epsilon^{(0)}_{k-q} - \epsilon^{(0)}_{p+q} + \epsilon^{(0)}_p)
\end{equation}
and the decay rate
\begin{equation}
    \frac{1}{\tau_k} = - 2\,{\rm Im} \Sigma^{(2)}(k,\omega)\big|_{\omega = \epsilon_k^{(0)}}.
\end{equation}
Near the pole, the dressed propagator $G_p^{d,(0)}(k,\omega)$ including the self-energy is given by
\begin{equation}
    G_p^{d,(0)}( k,\omega ) = \frac{1}{\omega - \epsilon_k - \Sigma^{(2)}(k, \omega)} \approx \frac{Z_k}{\omega - \tilde \epsilon_k + i \Gamma_k/2},
\end{equation}
with $\tilde{\epsilon}_k = \epsilon_k + {\rm Re} \Sigma^{(2)}$ and $\Gamma_k \equiv -2 {\rm Im} \Sigma^{(2)} \geq 0$. By performing the Fourier transformation in time, we obtain
\begin{equation}
    G_p^{d,(0)}( k,t) = \int \frac{d\omega}{2\pi} e^{-i\omega t} G_p^{d,(0)}( k,\omega )  = -i\theta(t) Z_k e^{-i\tilde \epsilon_k t} e^{-t/\tau_k}.
\end{equation}
This yields a modification to $G_p^{d,(0)}(x,t)$, contributing an additional exponential decay factor $e^{-t/\tau_k}$.

If the dispersion has an inflection point, both momentum and energy conservation can be satisfied, yielding a nonzero decay rate. For the non-interacting dispersion $\epsilon_k^{(0)}  = -2\cos k - \mu$, the on-shell condition
\begin{equation}
    \epsilon_k^{(0)} + \epsilon_p^{(0)} - \epsilon^{(0)}_{k-q} - \epsilon^{(0)}_{p+q} = 0
\end{equation}
is satisfied for
\begin{equation}
     p = \pi - k, \quad k + k_F - \pi < q < k - k_F.
\end{equation}
This decay process is allowed for $k > \pi - k_F$. It involves a finite-momentum scattering process in which the high-energy particle decays as a hole is created at momentum $\pi - k$, while another high-energy particle is created above the Fermi surface. There is a continuum of two-particle and one-hole excitations, corresponding to different choices of $q$, which are degenerate with the single-particle excitation. Importantly, the allowed range of $q$ shrinks to zero in the limit $k_F \rightarrow \pi/2$, and the decay rate vanishes at half-filling. 

The long-time behavior of the particle Green's function is then
\begin{equation}
    G_p(x, t\gg x/v, |m|x^2) \sim \frac{A e^{-i\epsilon_\pi t - t/\tau_\pi}}{t^{\eta_\pi}} + \frac{B}{t },
\end{equation}
where $B$ encodes Wen's result, which is robust against interactions, and $\eta_\pi = 1/2 + \nu_R$. Physically, the particle Green's function probes the spectrum above the chemical potential, as shown in Fig.~\ref{Appendix_Figure_1}(b1).

\subsection{Hole Green's function}
We are now ready to compute the time-dependent hole Green's function,
\begin{equation}
    G_h(x,t) = \langle \psi^\dagger(x,t)\psi(0,0) \rangle = G_h^R (x,t) + G_h^d(x,t),
\end{equation}
where $G_h^R(x,t) = \langle \psi^\dagger_R(x,t) \psi_R(0,0) \rangle$ and $G_h^d(x,t) = \langle d^\dagger(x,t) d(0,0) \rangle$. In the long-time limit $t \gg \epsilon^{-1}$, we can use the mode expansion and write
\begin{equation}
    G_h^d(k,t) \sim \int dx\, \langle d^\dagger(x,t) d(0,0) \rangle.
\end{equation}
In terms of the decoupled fields,
\begin{equation}
    \begin{aligned}
        G_h^d(k,t) \sim \int dx\, \langle \bar d^\dagger(x,t) \bar d(0,0) \rangle \,
        \langle e^{+i\sqrt{2\pi \nu_R}\bar \varphi_R(x,t)} e^{-i \sqrt{2\pi \nu_R} \bar \varphi_R(0,0)} \rangle,
    \end{aligned}
\end{equation}
where 
\begin{equation}
    \nu_R(k) = \frac{1}{K} \bigg( \frac{\gamma_{R}(k)}{2\pi} \bigg)^2
\end{equation}
is the anomalous exponent. The free propagator for the $d$ particle is 
\begin{equation}
    G_h^{d,(0)}(x,t) = \langle \bar d^\dagger(x,t) \bar d(0,0) \rangle = e^{+i\epsilon t}\delta(x - u t).
\end{equation}
The correlation functions of the free bosonic fields can be evaluated by standard methods, and we find
\begin{equation}
     G_h^d(k,t) \sim \frac{e^{-i{\mathcal E} t }}{t^{\eta_R}},
\end{equation}
with $\eta_R = \nu_R(k)$.

The long-time behavior of the fermionic Green's function in real space is dominated by the saddle-point contribution. For the hole Green's function, the relevant saddle corresponds to a hole at the bottom of the band, $k = 0$; for the particle Green's function, it corresponds to a particle at the top of the band, $k = \pi$, as discussed in the previous subsection. The high-energy excitations near these points have a parabolic dispersion, $\epsilon(k) \approx \epsilon + k^2/2m$, where $m < 0$ is the renormalized mass. Consequently, the propagator for the decoupled $\bar d$ particle reads (see Sec.~\ref{Supp_Sec:Mode_Expansion} for further details)
\begin{equation}\label{eq:supp:hole_bare_scaling}
    G_h^{d,(0)}(x,t) \sim \sqrt{\frac{m}{2\pi i t}} e^{+i\epsilon_0 t + i m x^2 /2t}.  
\end{equation}
Therefore, in the non-interacting case, this contribution (for the hole sector) oscillates at a high frequency $\epsilon_0 \sim O(1)$ for $t \gg |m|x^2$ and decays as $1/\sqrt{t}$. In the interacting case, the oscillation frequency becomes the renormalized energy of the single-particle excitation at the corresponding extremum (at $k=0$ for holes or $k=\pi$ for particles). In addition, the power-law exponent is modified by coupling to low-energy modes.

\begin{figure}[!h]
\centering 
\includegraphics[width=0.8\columnwidth]{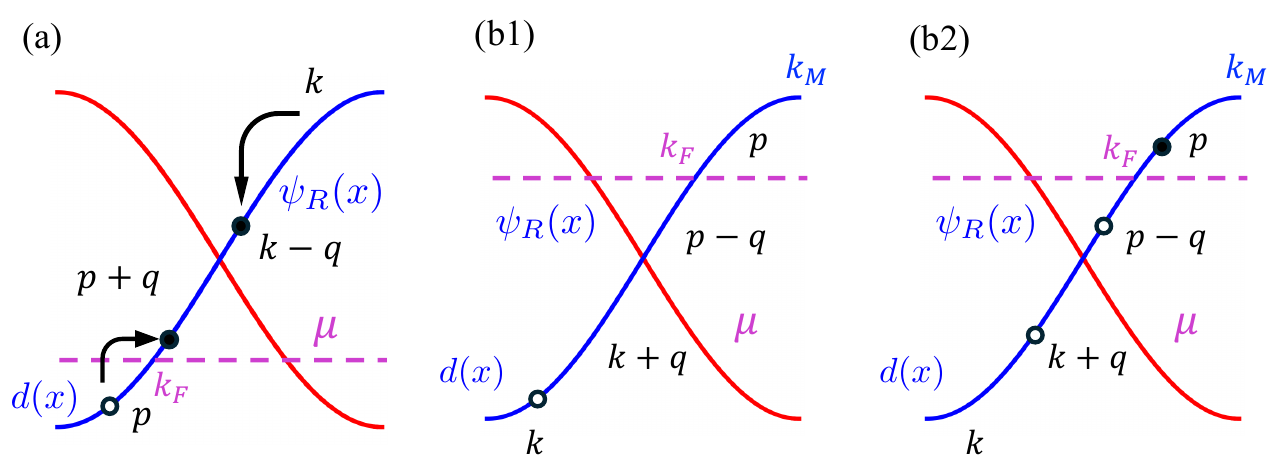}
\caption{\label{Appendix_Figure_1} Finite-momentum decay processes that generate a finite decay rate for high-energy excitations: (a) decay process contributing to the particle Green's function; (b) decay process contributing to the hole Green's function.}
\end{figure}

We consider a quasihole with momentum $k$ and energy $\omega$ below the Fermi level in the ground state $\ket{\rm GS}$, with the initial state $\ket{i} = \psi_k \ket{\rm GS}$ (see Fig.~\ref{Appendix_Figure_1}(b1)). To order $V^2$, the only inelastic decay channel is the creation of a particle-hole pair in the sea, with the final state $\ket{f} = \psi^\dagger_{p}\psi_{p-q} \psi_{k+q} \ket{\rm GS}$, subject to the constraints $n_p = 0$, $n_{p-q} = 1$, and $n_{k+q} = 1$ (i.e., $\theta(\epsilon^{(0)}_p)\,\theta(-\epsilon^{(0)}_{p-q})\,\theta(-\epsilon^{(0)}_{k+q})$). Energy conservation for the transition gives
\[
E_f - E_i = -\epsilon^{(0)}_{k+q} - \epsilon^{(0)}_{p-q} + \epsilon^{(0)}_p + \omega.
\]
The matrix element $M_{fi} = \langle f | H_{\rm int} | i \rangle = \frac{1}{2L}[V(q) - V(k - p - q)]$ has two contributions (direct and exchange). Fermi's golden rule yields the retarded self-energy
\begin{equation}
    -\frac{1}{2\tau_k} = {\rm Im}\,\Sigma^{(2)}(k,\omega)
    = -\frac{\pi}{2}\,[V(q) - V(k - p - q)]^2
    \,\theta(\epsilon^{(0)}_p)\,\theta(-\epsilon^{(0)}_{p-q})\,\theta(-\epsilon^{(0)}_{k+q})
    \,\delta\!\big(-\epsilon^{(0)}_{k+q} - \epsilon^{(0)}_{p-q} + \epsilon^{(0)}_p + \omega\big),
\end{equation}
and the decay rate
\begin{equation}
    \frac{1}{\tau_k} = -2\,{\rm Im}\,\Sigma^{(2)}(k,\omega)\big|_{\omega = -\epsilon_k^{(0)}}.
\end{equation}
If the dispersion has an inflection point, both momentum and energy conservation can be satisfied, leading to a nonzero decay rate. For the non-interacting dispersion $\epsilon_k^{(0)} = -2\cos k - \mu$, the on-shell condition
\begin{equation}
    -\epsilon^{(0)}_{k+q} - \epsilon^{(0)}_{p-q} + \epsilon^{(0)}_p + \omega = 0
\end{equation}
is satisfied for
\begin{equation}
     p = \pi - k, \qquad k + k_F - \pi < q < k - k_F.
\end{equation}
This decay process is allowed for $k > \pi - k_F$. It involves a finite-momentum scattering process in which the high-energy {hole} decays: a hole is created at momentum $\pi - k$ and, simultaneously, a high-energy particle is excited above the Fermi surface. There is a continuum of two-particle-one-hole excitations, corresponding to different choices of $q$, which are degenerate with the single-hole excitation. Importantly, the allowed range of $q$ shrinks to zero in the limit $k_F \rightarrow \pi/2$, and the decay rate vanishes at half-filling.

Similar to the discussion of the particle Green's function, the long-time behavior of the hole Green's function is given by
\begin{equation}
    G_h(x, t\gg x/v, |m|x^2) \sim \frac{A e^{+i\epsilon_0 t - t/\tau_0}}{t^{\eta_0}} + \frac{B}{t },
\end{equation}
where $\frac{B}{t }$ comes from conventional chiral Luttinger liquid theory~\cite{Wen1990PRBchiral}, which is robust against interactions, and $\eta_0 = 1/2 + \nu_R$. Physically, the hole Green's function probes the spectrum below the chemical potential, as illustrated in Fig.~\ref{Appendix_Figure_1}(b2).

\section{Details of mode expansion}\label{Supp_Sec:Mode_Expansion}
In this section, we discuss the mode expansion that leads to Eq.~\ref{eq:supp:particle_bare_scaling} and Eq.~\ref{eq:supp:hole_bare_scaling} introduced in Sec.~\ref{Supp_Sec:chiLL}. For simplicity, we consider the hole Green's function as an example~\cite{pereira2008exact,PereiraPRB2009Spectral}.

\subsection{Linearized model}
We begin by focusing on $t>0$ and on the real-time {hole} Green's function, defined with the conventional prefactor $i$,
\begin{equation}
    G_h(x,t) \equiv i \langle \psi^\dagger(x,t) \psi(0,0) \rangle,
\end{equation}
which measures the amplitude for removing a particle at the origin and later re-inserting it at position $x$ and time $t$.

In the continuum limit, we expand the field operators in momentum space using our Fourier convention,
\begin{equation}
    \psi(x,t) = \int \frac{dk}{2\pi} c_k e^{i kx - i\epsilon(k)t}, \quad \psi^\dagger(x,t) = \int \frac{dk}{2\pi} c_k^\dagger e^{-ikx + i \epsilon(k)t}.
\end{equation}
Here $c_k$ annihilates a fermion with momentum $k$, and $\epsilon(k)$ is the single-particle dispersion. In the ground state, Wick's theorem together with translational invariance implies
$\langle c_k^\dagger c_{k^\prime} \rangle = (2\pi)\delta(k-k^\prime)\, n(k)$, where $n(k)$ is the occupation function (at $T=0$, $n(k)$ reduces to the step function set by the Fermi sea). Substituting this into the definition of $G_h$ yields the standard momentum-space representation
\begin{equation}\label{eq:Appendix:G_h_Simple}
    G_h(x,t) = i \int \frac{dk}{2\pi} n(k) e^{-ikx + i \epsilon(k)t}.
\end{equation}

To obtain a simple, universal form, we now linearize the dispersion near a point $k_0$ with nonzero group velocity,
\begin{equation}
    \epsilon(k) \simeq \epsilon_0 + u(k-k_0),
\end{equation}
with $u = \epsilon^{\prime}(k_0)$. Physically, this approximation isolates the chiral low-energy modes in the vicinity of $k_0$ that dominate long-distance/long-time dynamics. Introducing the shifted variable $q = k - k_0$ and keeping only the slowly varying envelope, we find
\begin{equation}
    G_h(x,t) \simeq i e^{+i\epsilon_0 t} e^{-ik_0x} \int \frac{dq}{2\pi} e^{-iqx + iuq t} = i e^{+i\epsilon_0 t} e^{-ik_0x} \delta(x-ut).
\end{equation}
The integral evaluates to a Dirac delta because the phase is linear in $q$; this expresses ballistic propagation of the excitation at velocity $u$ along the characteristic $x=ut$, with the overall plane wave $e^{-ik_0 x}$ tracking the fast oscillations set by the expansion point.

If one absorbs the plane wave $e^{-ik_0x}$ into the definition of the slowly varying field (i.e., moves to the comoving envelope description), the compact statement is
\begin{equation}
    G_h^{(0)}(x,t) = ie^{+i\epsilon_0t} \delta(x-ut),
\end{equation}
which makes explicit that, within the linearized theory, the hole signal is sharply localized on the light cone defined by the group velocity $u$ and accumulates only the trivial on-shell phase $e^{+i\epsilon_0 t}$.

\subsection{Saddle point expansion}
At the bottom of the band (the hole edge), the group velocity vanishes, so the linear approximation breaks down and the leading curvature must be retained. Keeping the quadratic term,
\begin{equation}
    \epsilon(k) = \epsilon_0 + \frac{k^2}{2m_0}, \quad m_0 > 0,
\end{equation}
and substituting this into Eq.~\ref{eq:Appendix:G_h_Simple}, we obtain a Gaussian (Fresnel-type) integral over momentum,
\begin{equation}
    G_h^{(0)}(x,t) = ie^{i\epsilon_0t} \int \frac{dk}{2\pi} \exp \bigg( -ikx + i \frac{t}{2m_0} k^2 \bigg).
\end{equation}
This integral can be evaluated by completing the square or, equivalently, by invoking the standard Fresnel integral $\int \frac{dk}{2\pi} e^{i t k^2/(2m_0)}$, which yields the characteristic $t^{-1/2}$ decay and a dispersive phase. The result is
\begin{equation}
    G_h^{(0)}(x,t) = i \sqrt{\frac{m_0}{2\pi i t}} e^{i \epsilon_0 t} e^{-i \frac{m_0x^2}{2t}},
\end{equation}
where the square-root prefactor arises from the curvature-controlled stationary-phase evaluation near the saddle (located parametrically at $k^\ast \sim m_0 x/t$), and the additional exponential encodes the dispersive phase accumulated during propagation.

In the {time-dominated} window $t \gg m_0 x^2$, the phase $e^{-i m_0 x^2/(2t)}$ is slowly varying and can be expanded as $e^{-i m_0x^2/(2t)} = 1 + {\mathcal O}(x^2/t)$. Consequently, the free long-time asymptotics simplify to the universal power law
\begin{equation}
    G_h^{(0)}(x,t) \sim \frac{\tilde A_0 e^{i \epsilon_0 t}}{\sqrt{t}}.
\end{equation}
Here $\tilde A_0$ absorbs constant (model-dependent) phases and normalization factors, while the $t^{-1/2}$ decay reflects the dominance of the quadratic saddle at the band edge.

\subsection{Turning on interactions}
We now couple the high-energy {hole} to the gapless Luttinger mode that encodes long wavelength density fluctuations. Physically, this forward scattering coupling dresses the bare hole operator by a vertex operator of the bosonic field. As a result, its correlator acquires a universal power law factor governed by the orthogonality-catastrophe-type exponent. Concretely, for times well beyond the light-cone scale set by the Luttinger mode, i.e., in the regime $t \gg x/v$, the bosonic sector contributes a multiplicative decay $t^{-\Delta_h}$, where $\Delta_h$ is set by the interaction strength $V$. The total hole exponent is then
\begin{equation}
    \eta = \frac{1}{2} + \Delta_h,
\end{equation}
where the first term reflects the band-edge saddle and $\Delta_h$ encodes the interaction-induced dressing.

Combining this with the band-edge saddle (which supplies the $1/\sqrt{t}$ factor) yields the long-time asymptotics in the window $t \gg m_0 x^2$:
\begin{equation}
    G_h(x,t\gg x/v, m_0x^2) \sim \frac{A_h e^{+i \epsilon_0 t}}{t^{\eta}}.
\end{equation}
Here, $A_h$ is a nonuniversal amplitude absorbing short-distance (UV) details and overlap factors, while the on-shell phase $e^{+i\epsilon_0 t}$ reflects the band-edge energy of the dressed hole excitation.

\subsection{Two saddle points}
If the dispersion $\epsilon(k)$ is modified so that it develops two local minima (say at $k_0$ and $k_1$, with energies $\epsilon_0$ and $\epsilon_1$), the group velocity vanishes at both points and the leading contributions to the real-time propagator arise from {two} stationary points. In this situation, a stationary-phase (saddle-point) analysis of the integral representation--rather than a discrete sum over modes--naturally yields the superposition of the two saddle contributions. Starting from
\begin{equation}
\begin{aligned}
    G_h^{(0)}(x,t) =  \langle \bar d^\dagger(x,t) \bar d(0,0) \rangle  
    &= ie^{i\epsilon_0t} \int \frac{dk}{2\pi} \exp \bigg( -ikx + i \frac{t}{2m_0} k^2 \bigg) \\
    &\approx ie^{i\epsilon_0t} \int_{k\sim k_0} \frac{dk}{2\pi} \exp \bigg( -ikx + i \frac{t}{2m_0} k^2 \bigg) 
    + ie^{i\epsilon_0t} \int_{k \sim k_1} \frac{dk}{2\pi} \exp \bigg( -ikx + i \frac{t}{2m_0} k^2 \bigg)  \\  
    &\sim \sum_{j = 0 ,1} \sqrt{\frac{m_j}{2\pi i t}} e^{i \epsilon_j t} e^{-i m_j x^2/(2t)}, \quad m_j = [\epsilon^{\prime\prime}(k_j)]^{-1}>0,
\end{aligned}
\end{equation}
each minimum contributes a Gaussian saddle characterized by its local curvature $m_j = [\epsilon''(k_j)]^{-1}>0$, producing the familiar $t^{-1/2}$ prefactor and a dispersive phase $e^{-i m_j x^2/(2t)}$. The phases $e^{i\epsilon_j t}$ encode the distinct band-edge energies at the two minima. In the {time-dominated} regime $t \gg m_j x^2$ (for both $j=0,1$), these contributions interfere, leading to temporal beats at frequency $|\epsilon_1-\epsilon_0|$ and a spatially slowly varying envelope governed by $x^2/t$.

After dressing with interactions (Luttinger-mode coupling), each saddle picks up a universal power-law factor from the vertex-operator correlator and may acquire an additional exponential damping due to the finite lifetime of the high-energy excitation. Collecting these effects, we obtain
\begin{equation}
    G_h(x,t) \sim \frac{e^{-t/\tau}}{t^\eta} \bigg[ A_0 e^{i\epsilon_0 t} + A_1 e^{+i\epsilon_1t}\bigg],
\end{equation}
where $A_{0,1}$ are nonuniversal amplitudes (encoding overlaps and short-distance phases), $\eta$ is the interaction-dependent exponent, and $\tau$ is the lifetime. Both $\eta$ and $\tau$ are set by the interaction strength. This expression makes explicit the coexistence of two band-edge contributions and their mutual interference in the long-time dynamics.

\subsection{Final results}

In summary, for free fermions with two saddle points, a concise fit capturing the band-edge oscillation(s) and the universal tail is
\begin{equation}
    G_{p,h}(t) \sim \bigg( \frac{ A_0  + A_1 e^{-i\epsilon_{\rm Num}^{A_1}t}}{t^{\eta_{\rm Num}(0.5)}} + \frac{B_1e^{-i\epsilon_{\rm Num}^{B_1}t}}{t^{\alpha_{\rm Num}(1)}} \bigg).
\end{equation}
Here $A_0,A_1,B_1$ are amplitudes; $\epsilon_{\rm Num}^{A_1}$ sets the main band-edge frequency (energy scale between the chemical potential and band bottom of the edge dispersion), while $\epsilon_{\rm Num}^{B_1}$ captures a smaller edge scale (energy scale for the difference between the two saddle points). This form describes exactly the behavior in the checkerboard model of the main text, see Fig.~\ref{Figure_Free_Dynamic}(b), where a small and a large frequency oscillation are superimposed.

\section{Density-density correlator \label{sec:NN_corr}}
For chiral Luttinger liquids, we consider $N$ chiral bosons $\varphi(x,t) = (\varphi_1,\cdots, \varphi_N)$ governed by the multicomponent $K$-matrix action
\begin{equation}
    S = \frac{1}{4\pi} \int [dtdx] \,\big(\partial_t \varphi ^T K \partial_x \varphi - \partial_x \varphi^T U  \partial_x \varphi\big),
\end{equation}
where $K$ is an integer, symmetric, non-singular matrix encoding the topological data, and $U$ is a real, symmetric, positive-definite velocity/interactions matrix. The edge charge vector $t$ specifies how the chiral modes couple to the electromagnetic $U(1)$, so the physical edge charge density is
\[
n_x = \frac{1}{2\pi} t^T \partial_x \varphi(x).
\]
We are interested in the connected correlator
\begin{equation}\label{eq:Appendix:Connect_Correlation_Function}
    C(x) \equiv \langle n_x n_0 \rangle - \langle n_x \rangle \langle n_0 \rangle,
\end{equation}
evaluated at equal time $t=0$ (the subtraction removes the disconnected, uniform piece in a translationally invariant state).

The equal-time two-point function for chiral bosons, regularized with a short-distance cutoff $a>0$ (e.g., the lattice spacing), is
\begin{equation}
    \langle \varphi_I(x) \varphi_J(0) \rangle =  - (K^{-1})_{IJ} \ln \frac{x+ia}{a}.
\end{equation}
Noting that $n_0 = \frac{1}{2\pi}\, t^T \partial_y\varphi(y)\big|_{y = 0}$ and differentiating the logarithm in the distributional sense, Eq.~\eqref{eq:Appendix:Connect_Correlation_Function} becomes
\begin{equation}
    \begin{aligned}
       C(x) &\sim \frac{1}{4\pi^2}\,
       t^T \partial_x \partial_y \langle \varphi(x) \varphi(y) \rangle \big|_{y =0 } \\
       &= \frac{1}{4\pi^2}\,
       t^T \partial_x \partial_y \bigg[ -K^{-1} \ln \frac{x-y+ia}{a} \bigg] t \bigg|_{y = 0}
       \sim \frac{\nu}{4\pi^2}\, \frac{1}{(x+ia)^2},
    \end{aligned}
\end{equation}
where $\nu \equiv t^T K^{-1} t$ is the filling factor. Thus the density-density correlations decay as $1/x^2$ with a universal coefficient fixed solely by the topological data. For the $\nu=1/2$ case of interest, $\varphi(x,t)$ has one component with $t=1$ and $K=2$ (so $\nu = 1/2$), and the same formula applies with these identifications.

\begin{figure}[!h]
\centering 
\includegraphics[width=0.75\columnwidth]{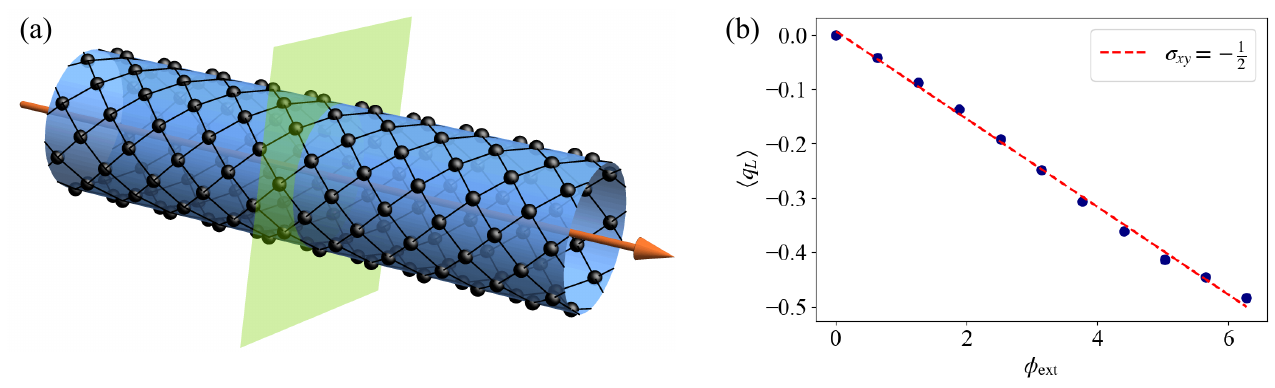}
\caption{\label{Appendix_Figure_2}
Static properties of the bosonic $\nu = 1/2$ fractional Chern insulator of the model Hamiltonian Eq.~(4) in the main text on an infinite cylinder. We insert a flux, as illustrated in (a), and monitor the charge on the left half of the system, shown in (b). The response shows the pumping of half a charge per $2\pi$ flux period, as expected for a $\nu = 1/2$ FCI state.
}
\end{figure}

\section{Further numerical characterization of the FCI }

Before presenting the numerical data, we briefly comment on why the static edge correlators agree more closely Wen's hydrodynamic prediction than the onsite dynamical Green's functions in the presence of a crystalline lattice. Band curvature and other lattice effects primarily distort the real-time propagation of high-energy edge modes, which strongly affects local-in-space correlators such as $G_h(t) = \langle a^\dagger_0(t) a_0(0)\rangle$. By contrast, equal-time correlators at large separations, such as $\langle a_0^\dagger a_x \rangle$ and the connected density-density correlator $C(x)$, are dominated by the low-energy chiral modes with approximately linear dispersion. As a result, their long-distance scaling more faithfully reflects the universal chiral Luttinger liquid exponents, even in a microscopic lattice model with substantial band curvature.

In this section, we show additional data from the matrix-product state (MPS) simulations characterizing the FCI state. First, we investigate the system under periodic boundary conditions in $y$ and particle filling $\nu_b=1/4$, i.e., half filling of the lowest band, and perform a flux insertion. The quantized Hall response clearly confirms the presence of the FCI.

Next, we show the static correlator $\langle a_{0}^{\dagger} a_{x}\rangle$ for different fillings. According to $\chi$LL theory, it should decay as $\propto 1/x^{1/\nu}$. This behavior is clearly present for all fillings at the highest bond dimension $\chi=3200$ except for $\nu_b=1/4$. In the case of exactly $\nu_b=1/4$, the correlator falls off faster at long distances. This is an indication that the excess charge density leads to a much stronger coupling between the opposite edges and that the boson density needs to be $\nu_b < 1/4$ in this finite system with open boundaries in the $y$-direction to obtain a well-established FCI state.

\begin{figure}[!h]
\centering 
\includegraphics[width=1\columnwidth]{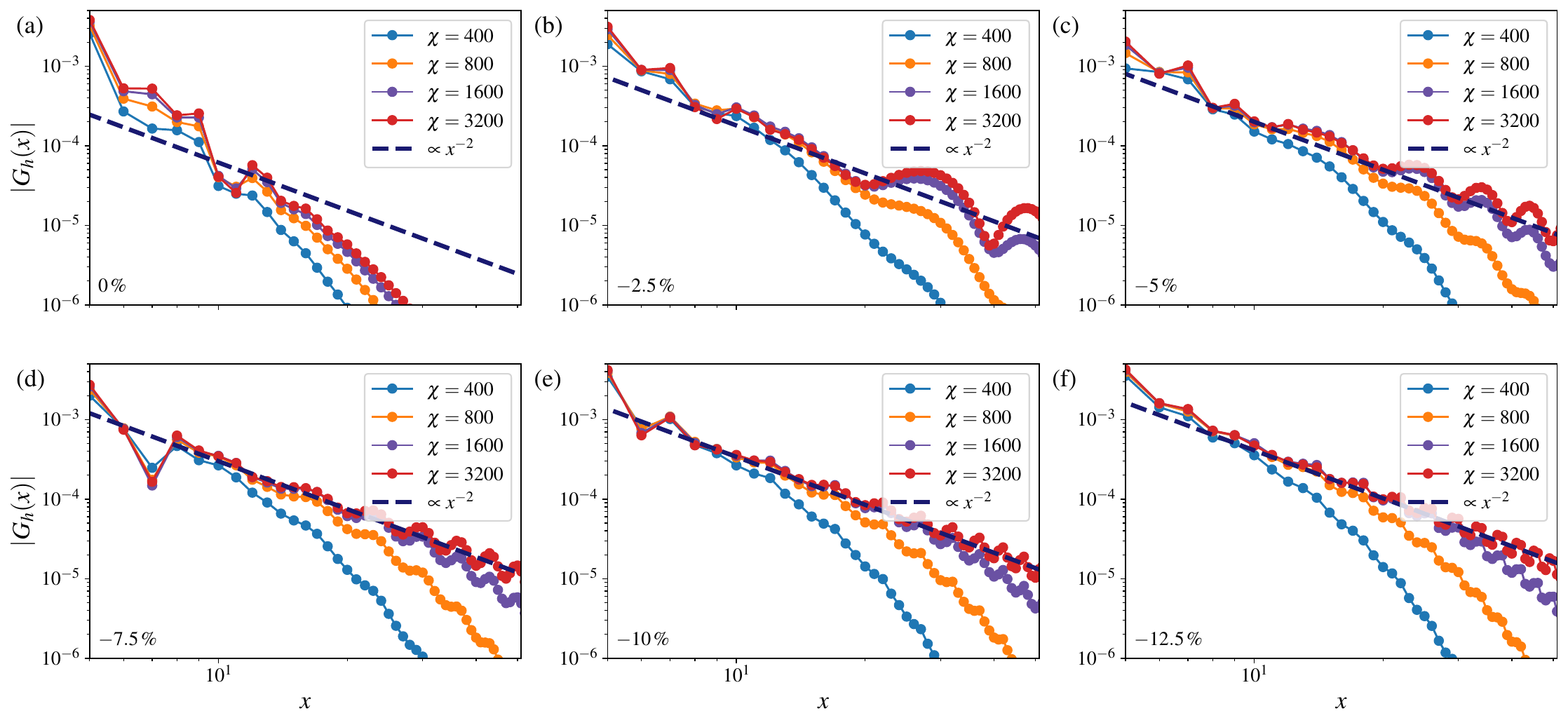}
\caption{\label{Appendix_Figure_3}
Static one-particle correlator $|G_h(x)| = |\langle a_0^\dagger a_x\rangle|$ at different fillings, computed along the lower edge for several bond dimensions $\chi = 400, 800, 1600, 3200$. Panels (a)-(f) correspond to deviations of the particle density from $\nu_b = 1/4$ by $0\%$, $-2.5\%$, $-7.5\%$, $-10\%$, $-12.5\%$ and $-15\%$, respectively. The dashed line indicates the $x^{-2}$ power-law behavior expected from chiral Luttinger liquid theory for a $\nu = 1/2$ FCI edge. As the filling is decreased below $\nu_b = 1/4$, the correlators at the largest bond dimension agree increasingly well with the predicted $1/x^2$ scaling, while at exactly $\nu_b = 1/4$ the correlator decays more rapidly due to stronger coupling between the opposite edges.
}
\end{figure}

\begin{figure}[!h]
\centering 
\includegraphics[width=1\columnwidth]{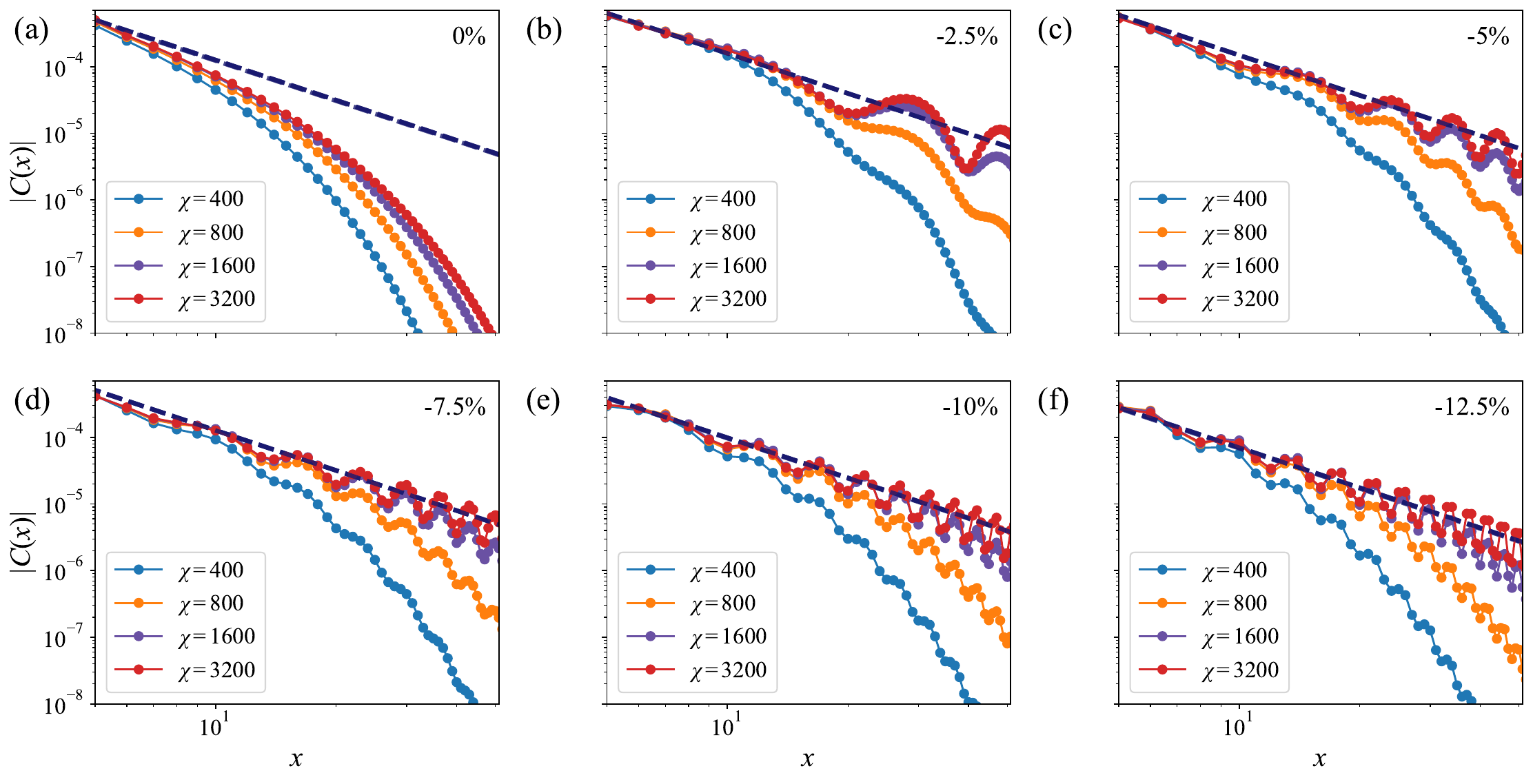}
\caption{\label{Appendix_Figure_4}
Connected density density correlator $|C(x)|$ at different fillings, computed along the lower edge for several bond dimensions $\chi = 400, 800, 1600, 3200$. Panels (a)-(f) correspond to deviations of the particle density from $\nu_b = 1/4$ by $0\%$, $-2.5\%$, $-5\%$, $-7.5\%$, $-10\%$, and $-12.5\%$, respectively. The dashed line indicates the $x^{-2}$ scaling expected from chiral Luttinger liquid theory for a $\nu = 1/2$ FCI edge. For all fillings below $\nu_b = 1/4$, the correlator at the largest bond dimension follows the predicted $1/x^2$ behavior over an extended range of distances, while at exactly $\nu_b = 1/4$ the correlations decay much faster, consistent with stronger coupling between the opposite edges in this finite strip geometry.
}
\end{figure}

The connected density-density correlator is shown in Fig.~\ref{Appendix_Figure_4}. As we showed in Sec.~\ref{sec:NN_corr}, it should decay as $1/x^2$ for the gapless edge state at any filling. This is confirmed by the data for all fillings smaller than $\nu_b=1/4$ that we consider here. As in the case of the $\langle a_{0}^{\dagger} a_{x}\rangle$ correlator, $|C(x)|$ again decays much faster for $\nu_b=1/4$.

\begin{figure}[!h]
\centering 
\includegraphics[width=1\columnwidth]{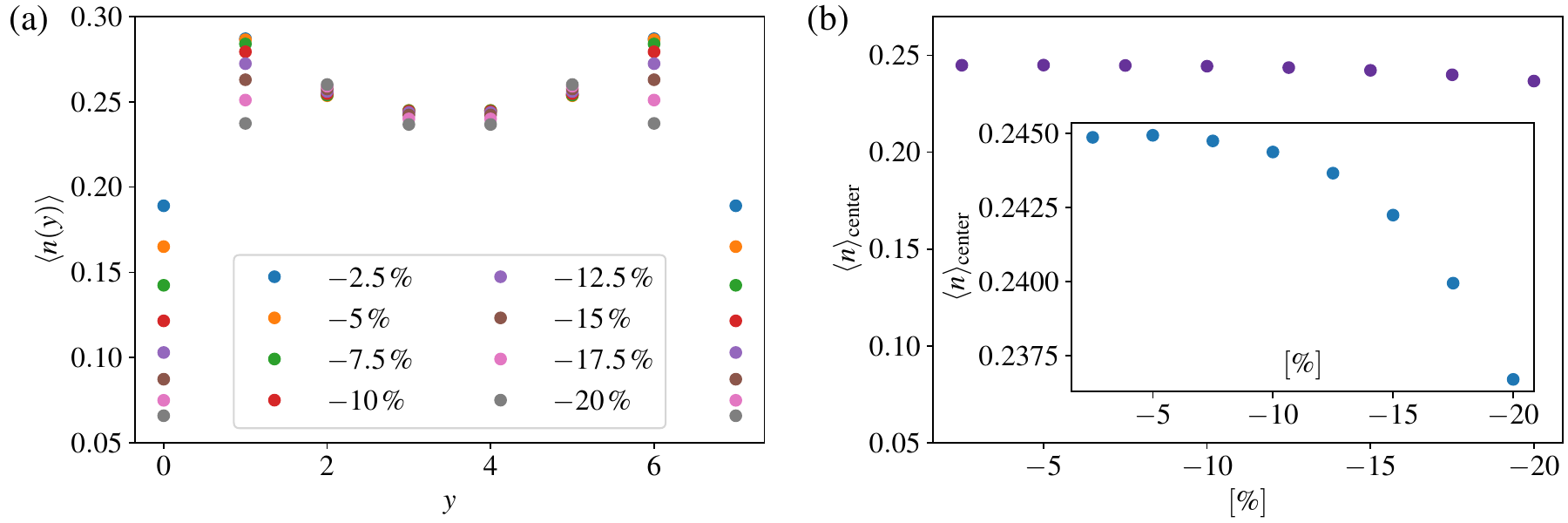}
\caption{\label{Appendix_Figure_Dop}
Charge density profile along $y$ direction for different fillings at bond dimension $\chi = 1600$, where the ground states have converged. 
(a) Layer-resolved density $\langle n(y)\rangle$ across the strip for several deviations of the boson filling $\nu_b$ from $1/4$, indicated in the legend. The density in the central rows of the strip remains nearly unchanged over a wide range of fillings, while the main changes occur at the outermost sites as particles are removed. 
(b) Average density in the center of the strip, $\langle n \rangle_{\mathrm{center}}$, as a function of the deviation from $\nu_b = 1/4$ (with the same filling offsets as in panel (a)); the inset shows a zoom of the weak variation. The fact that the bulk density changes only weakly even down to about $-15\%$ demonstrates the incompressible nature of the FCI state.
}
\end{figure}

\begin{figure}[!h]
\centering 
\includegraphics[width=1\columnwidth]{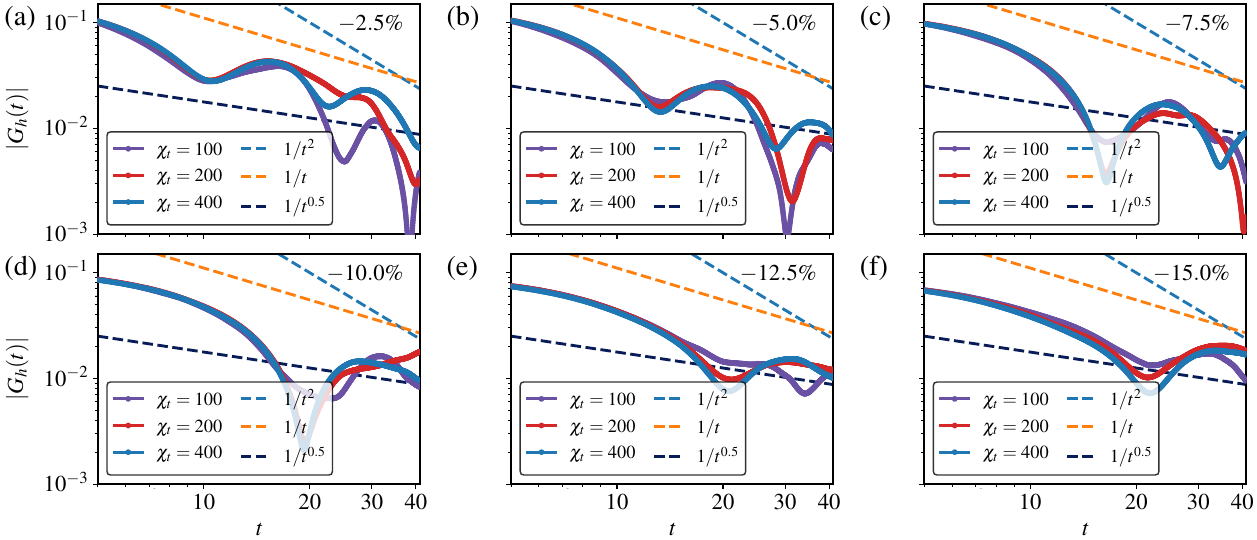}
\caption{\label{Appendix_Figure_5}
Dynamical onsite hole correlator $|G_h(t)| = |\langle b_0^\dagger(t) b_0(0)\rangle|$ for varying fillings and different MPS bond dimensions $\chi_t = 100, 200, 400$. Panels (a)-(f) correspond to deviations of the particle density from $\nu_b = 1/4$ by $-2.5\%$, $-5.0\%$, $-7.5\%$, $-10.0\%$, $-12.5\%$, and $-15.0\%$, respectively. The dashed lines show reference power laws $1/t^2$, $1/t$, and $1/t^{0.5}$, illustrating the expected crossover from the hydrodynamic $t^{-2}$ behavior toward a $t^{-1}$ decay modified by lattice effects. For all fillings, the curves for $\chi_t = 200$ and $\chi_t = 400$ agree well over the time window accessible in our simulations, indicating good convergence of the time evolution. The oscillation frequency of $|G_h(t)|$ changes monotonically with filling, consistent with the analytical parton picture discussed in the main text.
}
\end{figure}

Finally, we present the charge density $\langle n_y \rangle$ in Fig.~\ref{Appendix_Figure_Dop}. It stays approximately constant in the center of the strip over a large range of deviations from $\nu_b=1/4$, reflecting the incompressible nature of the FCI state. The density mainly changes at the two outermost sites when particles are removed or added.

In Fig.~\ref{Appendix_Figure_5}, we show the hole Green's function $\langle b^\dagger_0 (t) b_0(0)\rangle$ considered in the main text for different particle densities and bond dimensions $\chi=100,200,400$. The relatively low bond dimensions are due to the computational complexity of the time evolution, which far exceeds that of the ground-state DMRG calculations, and the need for a large infinite MPS unit cell to simulate the propagation of the perturbation. All calculations have been performed using the $W^{II}$ method from Ref.~\cite{Zaletel2015}. The different bond dimensions agree well for shorter times, but then deviate from one another. However, we already observe relatively good agreement between $\chi=200$ and $400$ in the time window we consider. For all fillings, we observe characteristic oscillations that are consistent with the analytical reasoning of the parton construction in analogy to the integer Chern insulator. Furthermore, the oscillations change frequency monotonically with filling, which is also in accordance with our analytical arguments.

\end{document}